\documentclass[letterpaper,english,reprint,aps,notitlepage,superscriptaddress,twocolumn]{revtex4-2}

\usepackage[T1]{fontenc}
\usepackage[utf8]{inputenc}
\usepackage{amsmath}
\usepackage{amssymb}
\usepackage[dvipsnames]{xcolor}

\colorlet{mylinkcolor}{RoyalPurple}
\colorlet{mycitecolor}{RoyalPurple}
\colorlet{myurlcolor}{RoyalPurple}

\usepackage{hyperref}
\hypersetup{
	linkcolor  = mylinkcolor,
	citecolor  = mycitecolor,
	urlcolor   = myurlcolor,
	colorlinks = true,
	breaklinks = true
}

\usepackage{graphicx}
\usepackage{babel}
\usepackage{physics}
\usepackage{tikz}
\usepackage{scalerel}
\usepackage{nicefrac}
\usepackage{xcolor}
\definecolor{mymagenta}{cmyk}{0,1,0,0.12}

\makeatletter

\pdfpageheight\paperheight
\pdfpagewidth\paperwidth

\usetikzlibrary{external}
\definecolor{primary}{RGB}{75,82,134}
\definecolor{accent}{RGB}{224,64,251}

\makeatother

\begin{document}

\title{Beating the spectroscopic Rayleigh limit via post-processed heterodyne detection}

\author{Wiktor Krokosz}
\affiliation{Centre for Quantum Optical Technologies, Centre of New Technologies, University of Warsaw, Banacha 2c, 02-097 Warsaw, Poland}
\affiliation{Faculty of Physics, University of Warsaw, Pasteura 5, 02-093 Warsaw, Poland}

\author{Mateusz Mazelanik}
\affiliation{Centre for Quantum Optical Technologies, Centre of New Technologies, University of Warsaw, Banacha 2c, 02-097 Warsaw, Poland}
%\affiliation{Faculty of Physics, University of Warsaw, Pasteura 5, 02-093 Warsaw, Poland}

\author{Michał Lipka}
\affiliation{Centre for Quantum Optical Technologies, Centre of New Technologies, University of Warsaw, Banacha 2c, 02-097 Warsaw, Poland}
%\affiliation{Faculty of Physics, University of Warsaw, Pasteura 5, 02-093 Warsaw, Poland}

\author{Marcin Jarzyna}
\affiliation{Centre for Quantum Optical Technologies, Centre of New Technologies, University of Warsaw, Banacha 2c, 02-097 Warsaw, Poland}
%\affiliation{Faculty of Physics, University of Warsaw, Pasteura 5, 02-093 Warsaw, Poland}
\author{Wojciech Wasilewski}
\affiliation{Centre for Quantum Optical Technologies, Centre of New Technologies, University of Warsaw, Banacha 2c, 02-097 Warsaw, Poland}
\affiliation{Faculty of Physics, University of Warsaw, Pasteura 5, 02-093 Warsaw, Poland}

\author{Konrad Banaszek}
\affiliation{Centre for Quantum Optical Technologies, Centre of New Technologies, University of Warsaw, Banacha 2c, 02-097 Warsaw, Poland}
\affiliation{Faculty of Physics, University of Warsaw, Pasteura 5, 02-093 Warsaw, Poland}

\author{Michał Parniak}
\email{m.parniak@cent.uw.edu.pl}
\affiliation{Centre for Quantum Optical Technologies, Centre of New Technologies, University of Warsaw, Banacha 2c, 02-097 Warsaw, Poland}

\begin{abstract}
Quantum-inspired superresolution methods surpass the Rayleigh limit in imaging, or the analogous Fourier limit in spectroscopy. This is achieved by carefully extracting the information carried in the emitted optical field by engineered measurements. An alternative to complex experimental setups is to use simple homodyne detection and customized data analysis.
We experimentally investigate this method in the time-frequency domain and demonstrate the spectroscopic superresolution for two distinct types of light sources: thermal and phase-averaged coherent states. The experimental results are backed by theoretical predictions based on estimation theory.
\end{abstract}
\maketitle

\section{Introduction}

Spectroscopic measurements serve as an indispensable tool for studying the properties of matter and light in a wide range of scientific disciplines, such as physics, chemistry \cite{Bec2020}, astronomy \cite{Kitchin1995}, biology \cite{Sahu2016}, or metrology \cite{Levine1999}. However, the achievable resolution of state-of-the-art spectrometers, including grating-based and Fourier spectrometers, is fundamentally constrained by the Fourier limit. This is especially important for emissive spectroscopy, where the properties of the light source can not be modified. In such scenario, when one deals with complex spectral distributions (e.g. multiple overlapping spectral lines of different widths), this limit is hard to overcome. However, there are cases in which specific characteristics of the light sources under study can be exploited to tailor an optimal measurement and beat this restriction by achieving a sub-Fourier resolution \cite{Donohue2018, Ansari2021, Mazelanik2022, Boschetti2020}. This is analogous to a similar effect present in optical imaging, where a Rayleigh criterion, formulated in a modern approach using tools from estimation theory, limits the fundamental spatial resolution of an image \cite{Tsang2016, Zhou2019a}. The basic example is a system of two identical and mutually incoherent point sources separated by some unknown distance that is much smaller than the spatial spread of each source. Many approaches to measuring the separation with sub-Rayleigh precision have been proposed \cite{Nair2016a, Tsang2019, Grace2020, Rehacek2018, Shah2021} and experimentally demonstrated \cite{Nair2016, Paur2016, Yang2016,  Parniak2018, Frank2023, Wadood2021, Zhou2019, Pushkina2021, Tan2023}. Such methods utilize shot-to-shot coherence and focus on information encoded within the electromagnetic (EM) field of light rather than its measured intensity as in traditional techniques. However, all of them require intricate and specialized equipment crafted to specific experimental conditions, as they rely on projecting the field on a specific set of modes. Another recently proposed approach, which also yields superior results compared to traditional intensity-based methods, is to use homodyne or heterodyne sensing with tailored digital postprocessing \cite{Datta2020, Datta2021}. A prominent advantage of such technique is a significantly simplified measurement setup, as the complexity is shifted to the data analysis step.

In this study, we present an experimental demonstration of spectral superresolution accomplished via heterodyne sensing, utilizing previously proposed theoretical frameworks \cite{Datta2020, Datta2021, Yang2017}. It is important to note that our approach differs from \cite{Yang2016} as we do not utilize the shaping of the local oscillator beam. By leveraging the advantages of heterodyne sensing, our method provides a more accessible and practical means of achieving spectral superresolution in cases where the signal is not dominated by the shot noise, with applications in domains where fine spectral discrimination plays a pivotal role. Potential candidates are lasing structures \cite{Jaffrennou2010}, optomechanics \cite{Andrews2014, Bagci2014, Thomas2021} or quantum transducers (e.g. based on Rydberg atoms) \cite{Borowka2023, Kumar2023, Tu2022} and sensors \cite{Jing2020}. Our analysis method leverages the limitations of a typical approach, reminiscent of direct imaging, where only a power spectral density of the signal is used to analyze the spectral lines. At the same time, the experimental setup remains essentially the same, providing a prospect for immediate applications.

\section{Heterodyne superresolving estimation}
\label{sec:method}

\begin{figure}[t]
    \centering
    \includegraphics[width=0.92\linewidth]{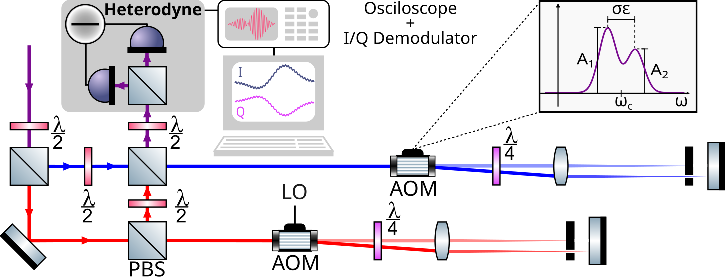}
    \caption{The experimental setup takes in a laser beam, which is then split in two using a half-wave plate $\left(\lambda/2\right)$ and a polarizing beam splitter (PBS). Both beams are then transmitted into one of the acousto-optic modulators (AOMs), which encode the local oscillator (LO) frequency as well as the programmatically set signal. The quarter-wave plate $\left(\lambda/4\right)$, convex lens, beam stop and mirror constitute a double pass AOM setup. Once the laser light is properly modulated, the two signals are then joined together and directed into a differential photodiode (DPD). Finally, the heterodyne signal is collected and I/Q demodulated.}
    \label{fig:schema}
\end{figure}

Our experimental demonstration is based on simulating and measuring the light source using an experimental setup presented in Fig.~\ref{fig:schema}. By employing an acousto-optic modulator (AOM) together with an arbitrary signal generator, we are able to encode the desired EM field envelope described below, which is later measured by the time-resolved heterodyne sensor. In this section we will describe our estimation approach. The detailed theoretical analysis of our method can be found in \cite{Datta2020, Datta2021}. 

Let us consider two spectral lines, each emitting light in a Gaussian temporal mode. The time envelope of such EM field, assuming the lines are separated by $\varepsilon\sigma$, is given by
\begin{equation}
    A(t+t_c) = G(t)  e^{i\omega_ct} \left(A_1e^{i \phi_1}e^{-i \frac{\sigma\varepsilon t}{2}} + A_2e^{-i \phi_2}e^{i \frac{\sigma\varepsilon t}{2}}\right), \label{eq:envelope-thermal}
\end{equation}
where $A_i,\,\phi_i$ for $i=1,2$ are respectively modulus value and phase of complex amplitude emitted by each line, $\omega_c$ and $t_c$ are displacements in frequency and time domains and $\sigma$ is the characteristic spread of the Gaussian temporal mode
\begin{equation}
     G(t) = \sqrt[4]{\frac{2\sigma^2}{\pi}} e^{-\sigma^2t^2}.
\end{equation}
We will specifically consider the cases of incoherent thermal radiation and coherent light averaged over the relative phase. In the former scenario, one has to use \eqref{eq:envelope-thermal} with amplitudes and phases $(A_i,\phi_i)$ randomly chosen in each realization from a complex Gaussian distribution for each line, that is, $A_i e^{i\phi_i} \sim \mathcal{CN}(0, \mathrm{A_0^2/2})$. The latter scenario, involves taking $A_1=A_2=A_0/\sqrt{2}$ and $\phi_1=\phi_2=\phi_0$ in \eqref{eq:envelope-thermal} with the phase $\phi_0$ randomly chosen from a flat distribution on the $2\pi$ period in each realization while $A_0$ is kept fixed i.e. only the relative phase is changing as the global phase referred to LO is unknown and changes in time.

Let us now consider the time-resolved heterodyne detection of an optical field described above. The analog heterodyne signal is digitalized and in-phase (I) and $\pi/2$-quadrature (Q) demodulated in real time and stored as complex signal traces $Z(t_i)=I(t_i)+iQ(t_i)$. Our estimation procedure begins by calculating the projections of the $Z(t_i)$ onto the first and second Hermite-Gauss (HG) modes:
\begin{equation}
    u^{(0)}(t) = G(t),\; u^{(1)}(t) = 2\sigma t G(t),
\end{equation}
where the Gaussian width $\sigma$ is set the same as in the input signal. Crucially, the second HG mode, in contrast to the first, is antisymmetric and therefore sensitive to discrepancies of the signal from $u^{(0)}(t)$, which describes a single line. For very small separations between the pulses, we can assume that almost all of the signal is distributed between these two modes \cite{Yang2017}. The projections are calculated using a simple discrete sum as
\begin{equation}
        z^{(k)}(t_r, \omega_r) = \sum_i Z(t_i)^* u^{(k)}(t_i - t_r) e^{i\omega_r t} \Delta t, \label{eq:gauss-projection}
\end{equation}
where $\omega_r,\,t_r$ are some spectral and temporal displacements.

Next, for each set of projections specified by the chosen values of $(\omega_r,t_r)$, we compute a variance $V^{(k)}=\mathrm{Var}[z^{(k)}(t_r, \omega_r)]$.
The variances represent the amount of signal in a given spatial mode, but also contain a shot-noise contribution and are in measurement units (e.g. Volts at photodetector). To recover the dimensionless signal contribution in the $u^{(1)}$ mode relative to the contribution in the fundamental $u^{(0)}$ mode, we calculate a minimal normalized and noise-subtracted variance:
\begin{equation}
    V_{\varepsilon} = \min_{t_r,\omega_r}\frac{V^{(1)} - V^{(1)}_{\mathrm{noise}}}{V^{(0)} - V^{(0)}_{\mathrm{noise}}},
    \label{eq:norm}
\end{equation}
where $V^{(k)}_{\mathrm{noise}}$ represents variance from a projection of a shot-noise calculated for measurement without the input signal. The values obtained from $t_r,\omega_r$ that minimize the variance serve as the centroid estimates that preserve high precision even for small separations \cite{Datta2020}. Finally, the variance $V_{\varepsilon}$ can be used to estimate the separation value:
\begin{equation}
    \hat{\varepsilon} = 4\sqrt{\max\{V_{\varepsilon}, 0\}}.
    \label{eq:estimator}
\end{equation}
Note that negative $V_\varepsilon$ values may arise when the signal contribution to $V^{(1)}$ is almost negligible. The estimation error may be found by employing the statistical bootstrapping method \cite{Efron1979}, described in Sec.~\ref{sec:results}.

Finally, to estimate the average number of registered photons we calculate the signal-to-noise ratio $\mathcal{S}$ by comparing the variances in the fundamental mode for signal and noise:
\begin{equation}
    \mathcal{S} = \frac{V^{(0)_{}} - V^{(0)}_{\mathrm{noise}}}{V^{(0)}_{\mathrm{noise}}},
    \label{eq:snr}
\end{equation}
for $(t_r,\omega_r)$ obtained from \eqref{eq:norm}. This approach is valid for small $\varepsilon$ and for $\varepsilon>1$ can be extended by including higher order modes in the formula. $\mathcal{S}$ is directly related to the number of photons $\bar{n}$ in the experiment by $\bar{n}=\mathcal{S}$. Notably, by following the above procedure, from a single measurement we estimate the centroids, separation, and average photon flux.

\section{Precision limits}
\label{sec:limits}

\begin{figure*}[t]
    \centering
    \includegraphics[width=\textwidth]{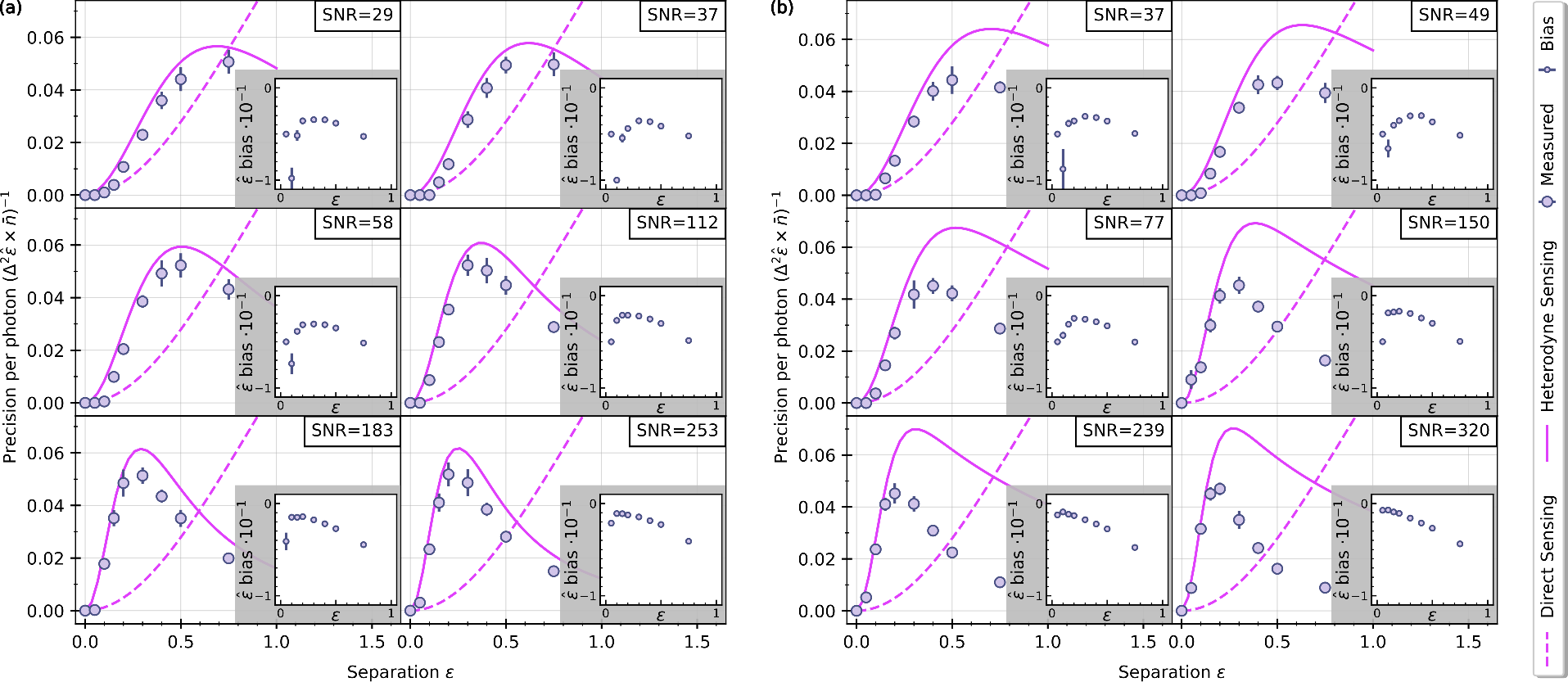}
    \caption{Plots representing the measured separation estimation precision in the case of thermal (a) and phase-averaged coherent (b) states. Each of the plots corresponds to a single $\mathcal{S}$ value, shown in ascending order. The data points are compared with the CRB given by the calculated FI for heterodyne sensing (solid curves) and DS (dashed curves). The error bars correspond to 2 standard deviations of the evaluated measured precision. The insets represent the estimator bias for each data point in the main plot.}
    \label{fig:fisher-eps}
\end{figure*}

The maximum precision limit of any estimation method based on a given measurement scheme is specified by the Cram\'er-Rao bound (CRB) \cite{Kay1993}. Assuming the outcomes $z$ are distributted according to a probability distribution $P(z|\varepsilon)$ the limit is equal to
\begin{equation}
(\Delta \varepsilon)^2\geq 1/F,\quad F=\int \frac{1}{P(z|\varepsilon)}\left[\frac{\partial P(z|\varepsilon)}{\partial \varepsilon}\right]^2\mathrm{d}z,\label{eq:CRB}
\end{equation}
where $F$ is Fisher information (FI).

The CRB for scenario described in the previous section Sec.~\ref{sec:method} can be derived from the measured quadratures statistic for both phase-averaged coherent and thermal states. However, direct calculation may be not be straightforward, particularly in the former case as the resulting quantum state is non-Gaussian \cite{Allevi2013, Olivares2012}. For this reason, we rather decompose the signal into HG modes and deal with each component separately.

The amplitude contribution from each HG mode can be calculated directly from the source spectral envelope as
\begin{equation}
    c_k(\varepsilon)=\int \mathrm{d}\omega \tilde{G}\left(\omega-\frac{\sigma \varepsilon}{2}\right)\tilde{u}^{(k)}(\omega)=\frac{e^{-\frac{\varepsilon^2}{32}}}{\sqrt{k!}}\left(\frac{\varepsilon}{4}\right)^k,
\end{equation}
for $k \in \{0,1\}$ and where $\tilde{G}$ and $\tilde{u}$ denote the time-domain Fourier transform of $G$ and $u$ respectively. Each source contributes with this amplitude to the measured signal in a given mode, and as a result, they interfere. For thermal sources, the result of the interference is still a thermal state with the mean photon number rescaled by $|c_k(\varepsilon)|^2$. Therefore, the quadrature distribution in the $u^{(1)}$ mode for the received state with $\bar{n}$ photons on average is given by:
\begin{equation}
    P_\mathrm{th}(z|\varepsilon)=\frac{1}{\pi (2|\alpha(\varepsilon)|^2+1)}\exp\left(-\frac{|z|^2}{2|\alpha(\varepsilon)|^2+1}\right), \label{eq:P_th}
\end{equation}
where $z$ is a complex variable and $\alpha(\varepsilon)=\sqrt{\bar{n}/2}c_{1}(\varepsilon)$ represents the average amplitude with $|\alpha(\varepsilon)|^2$ being average number of photons in the mode. The factor of $2$ that divides the number of photons is the result of using a heterodyne measurement \cite{Yang2017} in which each quadrature is measured on half of the signal and is accompanied by its individual shot noise.
In the case of phase-averaged coherent states, the output state is more complex with the quadrature distribution in the following form:
\begin{equation}
    P_{\mathrm{coh}}(z|\varepsilon) = \int_0^{2 \pi}\int_0^{2 \pi} \frac{\mathrm{d}\phi_1 \mathrm{d}\phi_2}{4\pi^3}  e^{-|z-\alpha(\varepsilon)(e^{i\phi_1}-e^{i\phi_2})|^2}.\label{eq:P_coh}
\end{equation}
The precision bounds may be found by plugging \eqref{eq:P_th} and \eqref{eq:P_coh} into the CRB in \eqref{eq:CRB}. Importantly, one often speaks about superresolution when the precision of separation estimation outperforms what can be achieved with direct sensing (DS). The bound for latter can be found by normalizing the modulus square of the EM field envelope and treating it as a probability distribution $P_{\mathrm{DS}}\sim |\tilde{A}(\omega)|^2$ for which one can evaluate the corresponding CRB.

Using the outcome distributions derived above, the per-photon FI for each mode can be calculated as:
\begin{equation}
    \mathcal{F}_{\mathrm{th/coh}}=\frac{1}{\bar{n}}F\left[P_{\mathrm{th/coh}}(z|\varepsilon)\right].\label{eq:FI_theory}
\end{equation}
For small separations ($\varepsilon\ll1$) the $u^\mathrm{(1)}$ mode contains almost all the information, and the other modes can be neglected. The formula for FI for thermal states can be evaluated analytically and yields:
\begin{equation}
    \mathcal{F}_{\mathrm{th}}=\frac{\bar{n}\left(\varepsilon/4\right)^2(\left(\varepsilon/4\right)^2-1)^2}{4\left(e^{\left(\varepsilon/4\right)^2}+\bar{n}\left(\varepsilon/4\right)^2\right)^2}.
    \label{eq:FI_th}
\end{equation}
In the case of phase-averaged coherent states the FI from \eqref{eq:FI_theory} is evaluated numerically to give the CRB for particular experimental conditions. The average number of photons $\bar{n}$ in the experiment is simply $\mathcal{S}$. 

\section{Results}
\label{sec:results}

For both thermal and phase-averaged coherent states, we generated Gaussian pulses with small spectral separations and measured them using the heterodyne detector. In the former case, the amplitudes and phases were sampled from a complex normal distribution, while in the latter, the amplitude was fixed and the phases were selected to uniformly cover the entire $2\pi$ period. To ensure the reliability and statistical significance of our results, we performed measurements for different mean signal intensities and varied the separations $\varepsilon$. For each parameter set, approximately $N\approx 10^6$ pulse repetitions were collected, providing a robust statistical sample. %The number of repetitions was chosen by observing the fluctuations in measured noise, which were sufficiently small for this order of magnitude.

%Note that this measurement and analysis procedure is similar to using a shaped LO beam \cite{Yang2016}, but here the shaping is done after the measurement.
Since we want to verify the saturability of the CRB, we need to calculate the variance of the estimator, given by \eqref{eq:estimator}. To do that, we employ the statistical bootstrapping method, where we generate 1000 ensembles of size $N$ from the calculated projections. Once we get the estimator values from each sample, we can calculate the precision by taking the inverse of the variance of the set multiplied by $\mathcal{S}$ and $N$.

The results obtained for the thermal states are shown in Fig. \ref{fig:fisher-eps}(a). We can clearly observe that the obtained precision taken as the inverse of the estimator's variance reliably follows the CRB (solid curve). For reference, we also show the CRB for DS (dashed line). From the plots, we see that the obtained precision beats the limit for DS proving the effectiveness of our method for achieving spectral superresolution in the case of thermal states.

For low $\mathcal{S}$, we observe the effect of noise that increased the estimator bias for the smallest separations \cite{Datta2020}. Additionally, for larger separations and $\mathcal{S}$, we observe that our estimator does not reach the CRB. We attribute this to fluctuations of the estimated $\mathcal{S}$, \eqref{eq:snr}, that also influence the normalized variance, \eqref{eq:norm}, used in the estimator. This is limited by the statistics from the shot-noise measurement, which could be improved. Currently, the number of shot-noise and signal measurements is equal, but one can increase the number of shot-noise measurements per experiment realization to estimate $\mathcal{S}$ more precisely. 
We can draw similar conclusions from the results obtained for phase-averaged coherent states depicted in Fig. \ref{fig:fisher-eps}(b). In this case the precision for $\varepsilon \ll 1$ follows CRB, and for larger separations we observe a larger decrease in precision. This is most likely a result of the non-optimal estimator that doesn't grasp the correlations between the two quadratures that become important for larger separations, as captured by the FI that is slightly larger when compared to FI for thermal states. 

\section{Summary}
In this study, we have experimentally demonstrated the heterodyne super-resolving measurement of two spectral lines for two different optical states: thermal and phase-averaged coherent. %Our experimental setup, which involved the use of an AOM and an arbitrary signal generator, allowed us to encode the desired EM field envelope, which was subsequently measured by a simple heterodyne sensor.
Our data analysis, based on numerical decomposition of measured quadrature distributions into first HG modes, enables estimating frequency separations and temporal and spectral centroid positions as well as average photon numbers from a single measurement set. %It is possible, to even extend the analysis to infer also the spectral widths of the sources, making the analysis free of nuisance parameters. 
The achieved precision of estimation was found to be in line with CRB derived for this measurement scheme, validating the effectiveness of our method for achieving spectral superresolution.
It is important to note that our method requires many photons per mode. While this may limit its applicability in certain scenarios, there are cases where this requirement can be met, and superresolution can be effectively achieved. These include optomechanical systems where mechanical modes are usually probed with a coherent light \cite{Tsaturyan2017} or lasing microstructures \cite{Jaffrennou2010}. 

In conclusion, our work contributes to ongoing efforts to overcome the fundamental constraints of state-of-the-art spectrometers, offering a more accessible and practical means of achieving spectral superresolution. This has potential applications in various scientific disciplines where fine spectral discrimination plays a pivotal role. In particular, our data analysis method may be readily applied in heterodyne spectroscopy setups which have so far used power spectral densities to recover properties of relevant spectral lines.

\section*{Data availability}
Data underlying the results presented in this paper are available in the Ref. \cite{dataset}.

\section*{Code availability}
The codes used for the calculations and the analysis of experimental data are available from the authors upon request.

%\section*{Author contributions}

\begin{acknowledgments}
The “Quantum Optical Technologies” (MAB/2018/4) project is carried out within the International Research Agendas programme of the Foundation for Polish Science co-financed by the European Union under the European Regional Development Fund.  This research was funded in whole or in part by the Office of Naval Research Global grant no. N62909-19-1-2127 and by National Science Centre, Poland 2021/41/N/ST2/02926. ML was supported by the Foundation for Polish Science (FNP) via the START scholarship
\end{acknowledgments}

\bibliographystyle{apsrev4-2}

\bibliography{biblio.bib}

\begin{thebibliography}{41}%
\makeatletter
\providecommand \@ifxundefined [1]{%
 \@ifx{#1\undefined}
}%
\providecommand \@ifnum [1]{%
 \ifnum #1\expandafter \@firstoftwo
 \else \expandafter \@secondoftwo
 \fi
}%
\providecommand \@ifx [1]{%
 \ifx #1\expandafter \@firstoftwo
 \else \expandafter \@secondoftwo
 \fi
}%
\providecommand \natexlab [1]{#1}%
\providecommand \enquote  [1]{``#1''}%
\providecommand \bibnamefont  [1]{#1}%
\providecommand \bibfnamefont [1]{#1}%
\providecommand \citenamefont [1]{#1}%
\providecommand \href@noop [0]{\@secondoftwo}%
\providecommand \href [0]{\begingroup \@sanitize@url \@href}%
\providecommand \@href[1]{\@@startlink{#1}\@@href}%
\providecommand \@@href[1]{\endgroup#1\@@endlink}%
\providecommand \@sanitize@url [0]{\catcode `\\12\catcode `\$12\catcode `\&12\catcode `\#12\catcode `\^12\catcode `\_12\catcode `\%12\relax}%
\providecommand \@@startlink[1]{}%
\providecommand \@@endlink[0]{}%
\providecommand \url  [0]{\begingroup\@sanitize@url \@url }%
\providecommand \@url [1]{\endgroup\@href {#1}{\urlprefix }}%
\providecommand \urlprefix  [0]{URL }%
\providecommand \Eprint [0]{\href }%
\providecommand \doibase [0]{https://doi.org/}%
\providecommand \selectlanguage [0]{\@gobble}%
\providecommand \bibinfo  [0]{\@secondoftwo}%
\providecommand \bibfield  [0]{\@secondoftwo}%
\providecommand \translation [1]{[#1]}%
\providecommand \BibitemOpen [0]{}%
\providecommand \bibitemStop [0]{}%
\providecommand \bibitemNoStop [0]{.\EOS\space}%
\providecommand \EOS [0]{\spacefactor3000\relax}%
\providecommand \BibitemShut  [1]{\csname bibitem#1\endcsname}%
\let\auto@bib@innerbib\@empty
%</preamble>
\bibitem [{\citenamefont {Beć}\ \emph {et~al.}(2020)\citenamefont {Beć}, \citenamefont {Grabska},\ and\ \citenamefont {Huck}}]{Bec2020}%
  \BibitemOpen
  \bibfield  {author} {\bibinfo {author} {\bibfnamefont {K.~B.}\ \bibnamefont {Beć}}, \bibinfo {author} {\bibfnamefont {J.}~\bibnamefont {Grabska}},\ and\ \bibinfo {author} {\bibfnamefont {C.~W.}\ \bibnamefont {Huck}},\ }\href {https://doi.org/10.1016/j.aca.2020.04.015} {\bibfield  {journal} {\bibinfo  {journal} {Analytica Chimica Acta}\ }\textbf {\bibinfo {volume} {1133}},\ \bibinfo {pages} {150} (\bibinfo {year} {2020})}\BibitemShut {NoStop}%
\bibitem [{\citenamefont {Kitchin}(1995)}]{Kitchin1995}%
  \BibitemOpen
  \bibfield  {author} {\bibinfo {author} {\bibfnamefont {C.~R.}\ \bibnamefont {Kitchin}},\ }\href@noop {} {\emph {\bibinfo {title} {Optical Astronomical Spectroscopy}}}\ (\bibinfo  {publisher} {Routledge \& CRC Press},\ \bibinfo {year} {1995})\BibitemShut {NoStop}%
\bibitem [{\citenamefont {Sahu}\ and\ \citenamefont {Mordechai}(2016)}]{Sahu2016}%
  \BibitemOpen
  \bibfield  {author} {\bibinfo {author} {\bibfnamefont {R.~K.}\ \bibnamefont {Sahu}}\ and\ \bibinfo {author} {\bibfnamefont {S.}~\bibnamefont {Mordechai}},\ }\href {https://doi.org/10.1080/05704928.2016.1157809} {\bibfield  {journal} {\bibinfo  {journal} {Applied Spectroscopy Reviews}\ }\textbf {\bibinfo {volume} {51}},\ \bibinfo {pages} {484} (\bibinfo {year} {2016})}\BibitemShut {NoStop}%
\bibitem [{\citenamefont {Levine}(1999)}]{Levine1999}%
  \BibitemOpen
  \bibfield  {author} {\bibinfo {author} {\bibfnamefont {J.}~\bibnamefont {Levine}},\ }\href {https://doi.org/10.1063/1.1149844} {\bibfield  {journal} {\bibinfo  {journal} {Review of Scientific Instruments}\ }\textbf {\bibinfo {volume} {70}},\ \bibinfo {pages} {2567} (\bibinfo {year} {1999})}\BibitemShut {NoStop}%
\bibitem [{\citenamefont {Donohue}\ \emph {et~al.}(2018)\citenamefont {Donohue}, \citenamefont {Ansari}, \citenamefont {\ifmmode \check{R}\else \v{R}\fi{}eh\'a\ifmmode~\check{c}\else \v{c}\fi{}ek}, \citenamefont {Hradil}, \citenamefont {Stoklasa}, \citenamefont {Pa\'ur}, \citenamefont {S\'anchez-Soto},\ and\ \citenamefont {Silberhorn}}]{Donohue2018}%
  \BibitemOpen
  \bibfield  {author} {\bibinfo {author} {\bibfnamefont {J.~M.}\ \bibnamefont {Donohue}}, \bibinfo {author} {\bibfnamefont {V.}~\bibnamefont {Ansari}}, \bibinfo {author} {\bibfnamefont {J.}~\bibnamefont {\ifmmode \check{R}\else \v{R}\fi{}eh\'a\ifmmode~\check{c}\else \v{c}\fi{}ek}}, \bibinfo {author} {\bibfnamefont {Z.}~\bibnamefont {Hradil}}, \bibinfo {author} {\bibfnamefont {B.}~\bibnamefont {Stoklasa}}, \bibinfo {author} {\bibfnamefont {M.}~\bibnamefont {Pa\'ur}}, \bibinfo {author} {\bibfnamefont {L.~L.}\ \bibnamefont {S\'anchez-Soto}},\ and\ \bibinfo {author} {\bibfnamefont {C.}~\bibnamefont {Silberhorn}},\ }\href {https://doi.org/10.1103/PhysRevLett.121.090501} {\bibfield  {journal} {\bibinfo  {journal} {Phys. Rev. Lett.}\ }\textbf {\bibinfo {volume} {121}},\ \bibinfo {pages} {090501} (\bibinfo {year} {2018})}\BibitemShut {NoStop}%
\bibitem [{\citenamefont {Ansari}\ \emph {et~al.}(2021)\citenamefont {Ansari}, \citenamefont {Brecht}, \citenamefont {Gil-Lopez}, \citenamefont {Donohue}, \citenamefont {Řeháček}, \citenamefont {Hradil}, \citenamefont {Sánchez-Soto},\ and\ \citenamefont {Silberhorn}}]{Ansari2021}%
  \BibitemOpen
  \bibfield  {author} {\bibinfo {author} {\bibfnamefont {V.}~\bibnamefont {Ansari}}, \bibinfo {author} {\bibfnamefont {B.}~\bibnamefont {Brecht}}, \bibinfo {author} {\bibfnamefont {J.}~\bibnamefont {Gil-Lopez}}, \bibinfo {author} {\bibfnamefont {J.~M.}\ \bibnamefont {Donohue}}, \bibinfo {author} {\bibfnamefont {J.}~\bibnamefont {Řeháček}}, \bibinfo {author} {\bibfnamefont {Z.}~\bibnamefont {Hradil}}, \bibinfo {author} {\bibfnamefont {L.~L.}\ \bibnamefont {Sánchez-Soto}},\ and\ \bibinfo {author} {\bibfnamefont {C.}~\bibnamefont {Silberhorn}},\ }\href {https://doi.org/10.1103/PRXQuantum.2.010301} {\bibfield  {journal} {\bibinfo  {journal} {PRX Quantum}\ }\textbf {\bibinfo {volume} {2}},\ \bibinfo {pages} {010301} (\bibinfo {year} {2021})}\BibitemShut {NoStop}%
\bibitem [{\citenamefont {Mazelanik}\ \emph {et~al.}(2022)\citenamefont {Mazelanik}, \citenamefont {Leszczy{\'{n}}ski},\ and\ \citenamefont {Parniak}}]{Mazelanik2022}%
  \BibitemOpen
  \bibfield  {author} {\bibinfo {author} {\bibfnamefont {M.}~\bibnamefont {Mazelanik}}, \bibinfo {author} {\bibfnamefont {A.}~\bibnamefont {Leszczy{\'{n}}ski}},\ and\ \bibinfo {author} {\bibfnamefont {M.}~\bibnamefont {Parniak}},\ }\href {https://doi.org/10.1038/s41467-022-28066-5} {\bibfield  {journal} {\bibinfo  {journal} {Nature Communications}\ }\textbf {\bibinfo {volume} {13}},\ \bibinfo {pages} {691} (\bibinfo {year} {2022})}\BibitemShut {NoStop}%
\bibitem [{\citenamefont {Boschetti}\ \emph {et~al.}(2020)\citenamefont {Boschetti}, \citenamefont {Taschin}, \citenamefont {Bartolini}, \citenamefont {Tiwari}, \citenamefont {Pattelli}, \citenamefont {Torre},\ and\ \citenamefont {Wiersma}}]{Boschetti2020}%
  \BibitemOpen
  \bibfield  {author} {\bibinfo {author} {\bibfnamefont {A.}~\bibnamefont {Boschetti}}, \bibinfo {author} {\bibfnamefont {A.}~\bibnamefont {Taschin}}, \bibinfo {author} {\bibfnamefont {P.}~\bibnamefont {Bartolini}}, \bibinfo {author} {\bibfnamefont {A.~K.}\ \bibnamefont {Tiwari}}, \bibinfo {author} {\bibfnamefont {L.}~\bibnamefont {Pattelli}}, \bibinfo {author} {\bibfnamefont {R.}~\bibnamefont {Torre}},\ and\ \bibinfo {author} {\bibfnamefont {D.~S.}\ \bibnamefont {Wiersma}},\ }\href {https://doi.org/10.1038/s41566-019-0558-4} {\bibfield  {journal} {\bibinfo  {journal} {Nature Photonics}\ }\textbf {\bibinfo {volume} {14}},\ \bibinfo {pages} {177} (\bibinfo {year} {2020})}\BibitemShut {NoStop}%
\bibitem [{\citenamefont {Tsang}\ \emph {et~al.}(2016)\citenamefont {Tsang}, \citenamefont {Nair},\ and\ \citenamefont {Lu}}]{Tsang2016}%
  \BibitemOpen
  \bibfield  {author} {\bibinfo {author} {\bibfnamefont {M.}~\bibnamefont {Tsang}}, \bibinfo {author} {\bibfnamefont {R.}~\bibnamefont {Nair}},\ and\ \bibinfo {author} {\bibfnamefont {X.-M.}\ \bibnamefont {Lu}},\ }\href {https://doi.org/10.1103/PhysRevX.6.031033} {\bibfield  {journal} {\bibinfo  {journal} {Phys. Rev. X}\ }\textbf {\bibinfo {volume} {6}},\ \bibinfo {pages} {031033} (\bibinfo {year} {2016})}\BibitemShut {NoStop}%
\bibitem [{\citenamefont {Zhou}\ and\ \citenamefont {Jiang}(2019)}]{Zhou2019a}%
  \BibitemOpen
  \bibfield  {author} {\bibinfo {author} {\bibfnamefont {S.}~\bibnamefont {Zhou}}\ and\ \bibinfo {author} {\bibfnamefont {L.}~\bibnamefont {Jiang}},\ }\href {https://doi.org/10.1103/PhysRevA.99.013808} {\bibfield  {journal} {\bibinfo  {journal} {Phys. Rev. A}\ }\textbf {\bibinfo {volume} {99}},\ \bibinfo {pages} {013808} (\bibinfo {year} {2019})}\BibitemShut {NoStop}%
\bibitem [{\citenamefont {Nair}\ and\ \citenamefont {Tsang}(2016{\natexlab{a}})}]{Nair2016a}%
  \BibitemOpen
  \bibfield  {author} {\bibinfo {author} {\bibfnamefont {R.}~\bibnamefont {Nair}}\ and\ \bibinfo {author} {\bibfnamefont {M.}~\bibnamefont {Tsang}},\ }\href {https://doi.org/10.1103/PhysRevLett.117.190801} {\bibfield  {journal} {\bibinfo  {journal} {Physical Review Letters}\ }\textbf {\bibinfo {volume} {117}},\ \bibinfo {pages} {190801} (\bibinfo {year} {2016}{\natexlab{a}})}\BibitemShut {NoStop}%
\bibitem [{\citenamefont {Tsang}(2019)}]{Tsang2019}%
  \BibitemOpen
  \bibfield  {author} {\bibinfo {author} {\bibfnamefont {M.}~\bibnamefont {Tsang}},\ }\href {https://doi.org/10.1080/00107514.2020.1736375} {\bibfield  {journal} {\bibinfo  {journal} {Contemporary Physics}\ }\textbf {\bibinfo {volume} {60}},\ \bibinfo {pages} {279} (\bibinfo {year} {2019})}\BibitemShut {NoStop}%
\bibitem [{\citenamefont {Grace}\ \emph {et~al.}(2020)\citenamefont {Grace}, \citenamefont {Dutton}, \citenamefont {Ashok},\ and\ \citenamefont {Guha}}]{Grace2020}%
  \BibitemOpen
  \bibfield  {author} {\bibinfo {author} {\bibfnamefont {M.~R.}\ \bibnamefont {Grace}}, \bibinfo {author} {\bibfnamefont {Z.}~\bibnamefont {Dutton}}, \bibinfo {author} {\bibfnamefont {A.}~\bibnamefont {Ashok}},\ and\ \bibinfo {author} {\bibfnamefont {S.}~\bibnamefont {Guha}},\ }\href {https://doi.org/10.1364/JOSAA.392116} {\bibfield  {journal} {\bibinfo  {journal} {J. Opt. Soc. Am. A}\ }\textbf {\bibinfo {volume} {37}},\ \bibinfo {pages} {1288} (\bibinfo {year} {2020})}\BibitemShut {NoStop}%
\bibitem [{\citenamefont {Řeháček}\ \emph {et~al.}(2018)\citenamefont {Řeháček}, \citenamefont {Hradil}, \citenamefont {Koutný}, \citenamefont {Grover}, \citenamefont {Krzic},\ and\ \citenamefont {Sánchez-Soto}}]{Rehacek2018}%
  \BibitemOpen
  \bibfield  {author} {\bibinfo {author} {\bibfnamefont {J.}~\bibnamefont {Řeháček}}, \bibinfo {author} {\bibfnamefont {Z.}~\bibnamefont {Hradil}}, \bibinfo {author} {\bibfnamefont {D.}~\bibnamefont {Koutný}}, \bibinfo {author} {\bibfnamefont {J.}~\bibnamefont {Grover}}, \bibinfo {author} {\bibfnamefont {A.}~\bibnamefont {Krzic}},\ and\ \bibinfo {author} {\bibfnamefont {L.~L.}\ \bibnamefont {Sánchez-Soto}},\ }\href {https://doi.org/10.1103/PhysRevA.98.012103} {\bibfield  {journal} {\bibinfo  {journal} {Physical Review A}\ }\textbf {\bibinfo {volume} {98}},\ \bibinfo {pages} {012103} (\bibinfo {year} {2018})}\BibitemShut {NoStop}%
\bibitem [{\citenamefont {Shah}\ and\ \citenamefont {Fan}(2021)}]{Shah2021}%
  \BibitemOpen
  \bibfield  {author} {\bibinfo {author} {\bibfnamefont {M.}~\bibnamefont {Shah}}\ and\ \bibinfo {author} {\bibfnamefont {L.}~\bibnamefont {Fan}},\ }\href {https://doi.org/10.1103/PhysRevApplied.15.034071} {\bibfield  {journal} {\bibinfo  {journal} {Physical Review Applied}\ }\textbf {\bibinfo {volume} {15}},\ \bibinfo {pages} {034071} (\bibinfo {year} {2021})}\BibitemShut {NoStop}%
\bibitem [{\citenamefont {Nair}\ and\ \citenamefont {Tsang}(2016{\natexlab{b}})}]{Nair2016}%
  \BibitemOpen
  \bibfield  {author} {\bibinfo {author} {\bibfnamefont {R.}~\bibnamefont {Nair}}\ and\ \bibinfo {author} {\bibfnamefont {M.}~\bibnamefont {Tsang}},\ }\href {https://doi.org/10.1364/OE.24.003684} {\bibfield  {journal} {\bibinfo  {journal} {Optics Express}\ }\textbf {\bibinfo {volume} {24}},\ \bibinfo {pages} {3684} (\bibinfo {year} {2016}{\natexlab{b}})}\BibitemShut {NoStop}%
\bibitem [{\citenamefont {Paúr}\ \emph {et~al.}(2016)\citenamefont {Paúr}, \citenamefont {Stoklasa}, \citenamefont {Hradil}, \citenamefont {Sánchez-Soto},\ and\ \citenamefont {Rehacek}}]{Paur2016}%
  \BibitemOpen
  \bibfield  {author} {\bibinfo {author} {\bibfnamefont {M.}~\bibnamefont {Paúr}}, \bibinfo {author} {\bibfnamefont {B.}~\bibnamefont {Stoklasa}}, \bibinfo {author} {\bibfnamefont {Z.}~\bibnamefont {Hradil}}, \bibinfo {author} {\bibfnamefont {L.~L.}\ \bibnamefont {Sánchez-Soto}},\ and\ \bibinfo {author} {\bibfnamefont {J.}~\bibnamefont {Rehacek}},\ }\href {https://doi.org/10.1364/OPTICA.3.001144} {\bibfield  {journal} {\bibinfo  {journal} {Optica}\ }\textbf {\bibinfo {volume} {3}},\ \bibinfo {pages} {1144} (\bibinfo {year} {2016})}\BibitemShut {NoStop}%
\bibitem [{\citenamefont {Yang}\ \emph {et~al.}(2016)\citenamefont {Yang}, \citenamefont {Tashchilina}, \citenamefont {Moiseev}, \citenamefont {Simon},\ and\ \citenamefont {Lvovsky}}]{Yang2016}%
  \BibitemOpen
  \bibfield  {author} {\bibinfo {author} {\bibfnamefont {F.}~\bibnamefont {Yang}}, \bibinfo {author} {\bibfnamefont {A.}~\bibnamefont {Tashchilina}}, \bibinfo {author} {\bibfnamefont {E.~S.}\ \bibnamefont {Moiseev}}, \bibinfo {author} {\bibfnamefont {C.}~\bibnamefont {Simon}},\ and\ \bibinfo {author} {\bibfnamefont {A.~I.}\ \bibnamefont {Lvovsky}},\ }\href {https://doi.org/10.1364/OPTICA.3.001148} {\bibfield  {journal} {\bibinfo  {journal} {Optica}\ }\textbf {\bibinfo {volume} {3}},\ \bibinfo {pages} {1148} (\bibinfo {year} {2016})}\BibitemShut {NoStop}%
\bibitem [{\citenamefont {Parniak}\ \emph {et~al.}(2018)\citenamefont {Parniak}, \citenamefont {Borówka}, \citenamefont {Boroszko}, \citenamefont {Wasilewski}, \citenamefont {Banaszek},\ and\ \citenamefont {Demkowicz-Dobrzański}}]{Parniak2018}%
  \BibitemOpen
  \bibfield  {author} {\bibinfo {author} {\bibfnamefont {M.}~\bibnamefont {Parniak}}, \bibinfo {author} {\bibfnamefont {S.}~\bibnamefont {Borówka}}, \bibinfo {author} {\bibfnamefont {K.}~\bibnamefont {Boroszko}}, \bibinfo {author} {\bibfnamefont {W.}~\bibnamefont {Wasilewski}}, \bibinfo {author} {\bibfnamefont {K.}~\bibnamefont {Banaszek}},\ and\ \bibinfo {author} {\bibfnamefont {R.}~\bibnamefont {Demkowicz-Dobrzański}},\ }\href {https://doi.org/10.1103/PhysRevLett.121.250503} {\bibfield  {journal} {\bibinfo  {journal} {Physical Review Letters}\ }\textbf {\bibinfo {volume} {121}},\ \bibinfo {pages} {250503} (\bibinfo {year} {2018})}\BibitemShut {NoStop}%
\bibitem [{\citenamefont {Frank}\ \emph {et~al.}(2023)\citenamefont {Frank}, \citenamefont {Duplinskiy}, \citenamefont {Bearne},\ and\ \citenamefont {Lvovsky}}]{Frank2023}%
  \BibitemOpen
  \bibfield  {author} {\bibinfo {author} {\bibfnamefont {J.}~\bibnamefont {Frank}}, \bibinfo {author} {\bibfnamefont {A.}~\bibnamefont {Duplinskiy}}, \bibinfo {author} {\bibfnamefont {K.}~\bibnamefont {Bearne}},\ and\ \bibinfo {author} {\bibfnamefont {A.~I.}\ \bibnamefont {Lvovsky}},\ }\href {https://doi.org/10.1364/OPTICA.493718} {\bibfield  {journal} {\bibinfo  {journal} {Optica}\ }\textbf {\bibinfo {volume} {10}},\ \bibinfo {pages} {1147} (\bibinfo {year} {2023})}\BibitemShut {NoStop}%
\bibitem [{\citenamefont {Wadood}\ \emph {et~al.}(2021)\citenamefont {Wadood}, \citenamefont {Liang}, \citenamefont {Zhou}, \citenamefont {Yang}, \citenamefont {Alonso}, \citenamefont {Qian}, \citenamefont {Malhotra}, \citenamefont {Rafsanjani}, \citenamefont {Jordan}, \citenamefont {Boyd},\ and\ \citenamefont {Vamivakas}}]{Wadood2021}%
  \BibitemOpen
  \bibfield  {author} {\bibinfo {author} {\bibfnamefont {S.~A.}\ \bibnamefont {Wadood}}, \bibinfo {author} {\bibfnamefont {K.}~\bibnamefont {Liang}}, \bibinfo {author} {\bibfnamefont {Y.}~\bibnamefont {Zhou}}, \bibinfo {author} {\bibfnamefont {J.}~\bibnamefont {Yang}}, \bibinfo {author} {\bibfnamefont {M.~A.}\ \bibnamefont {Alonso}}, \bibinfo {author} {\bibfnamefont {X.-F.}\ \bibnamefont {Qian}}, \bibinfo {author} {\bibfnamefont {T.}~\bibnamefont {Malhotra}}, \bibinfo {author} {\bibfnamefont {S.~M.~H.}\ \bibnamefont {Rafsanjani}}, \bibinfo {author} {\bibfnamefont {A.~N.}\ \bibnamefont {Jordan}}, \bibinfo {author} {\bibfnamefont {R.~W.}\ \bibnamefont {Boyd}},\ and\ \bibinfo {author} {\bibfnamefont {A.~N.}\ \bibnamefont {Vamivakas}},\ }\href {https://doi.org/10.1364/OE.427734} {\bibfield  {journal} {\bibinfo  {journal} {Optics Express}\ }\textbf {\bibinfo {volume} {29}},\ \bibinfo {pages} {22034} (\bibinfo {year} {2021})}\BibitemShut {NoStop}%
\bibitem [{\citenamefont {Zhou}\ \emph {et~al.}(2019)\citenamefont {Zhou}, \citenamefont {Yang}, \citenamefont {Hassett}, \citenamefont {Rafsanjani}, \citenamefont {Mirhosseini}, \citenamefont {Vamivakas}, \citenamefont {Jordan}, \citenamefont {Shi},\ and\ \citenamefont {Boyd}}]{Zhou2019}%
  \BibitemOpen
  \bibfield  {author} {\bibinfo {author} {\bibfnamefont {Y.}~\bibnamefont {Zhou}}, \bibinfo {author} {\bibfnamefont {J.}~\bibnamefont {Yang}}, \bibinfo {author} {\bibfnamefont {J.~D.}\ \bibnamefont {Hassett}}, \bibinfo {author} {\bibfnamefont {S.~M.~H.}\ \bibnamefont {Rafsanjani}}, \bibinfo {author} {\bibfnamefont {M.}~\bibnamefont {Mirhosseini}}, \bibinfo {author} {\bibfnamefont {A.~N.}\ \bibnamefont {Vamivakas}}, \bibinfo {author} {\bibfnamefont {A.~N.}\ \bibnamefont {Jordan}}, \bibinfo {author} {\bibfnamefont {Z.}~\bibnamefont {Shi}},\ and\ \bibinfo {author} {\bibfnamefont {R.~W.}\ \bibnamefont {Boyd}},\ }\href {https://doi.org/10.1364/OPTICA.6.000534} {\bibfield  {journal} {\bibinfo  {journal} {Optica}\ }\textbf {\bibinfo {volume} {6}},\ \bibinfo {pages} {534} (\bibinfo {year} {2019})}\BibitemShut {NoStop}%
\bibitem [{\citenamefont {Pushkina}\ \emph {et~al.}(2021)\citenamefont {Pushkina}, \citenamefont {Maltese}, \citenamefont {Costa-Filho}, \citenamefont {Patel},\ and\ \citenamefont {Lvovsky}}]{Pushkina2021}%
  \BibitemOpen
  \bibfield  {author} {\bibinfo {author} {\bibfnamefont {A.}~\bibnamefont {Pushkina}}, \bibinfo {author} {\bibfnamefont {G.}~\bibnamefont {Maltese}}, \bibinfo {author} {\bibfnamefont {J.}~\bibnamefont {Costa-Filho}}, \bibinfo {author} {\bibfnamefont {P.}~\bibnamefont {Patel}},\ and\ \bibinfo {author} {\bibfnamefont {A.}~\bibnamefont {Lvovsky}},\ }\href {https://doi.org/10.1103/PhysRevLett.127.253602} {\bibfield  {journal} {\bibinfo  {journal} {Physical Review Letters}\ }\textbf {\bibinfo {volume} {127}},\ \bibinfo {pages} {253602} (\bibinfo {year} {2021})}\BibitemShut {NoStop}%
\bibitem [{\citenamefont {Tan}\ \emph {et~al.}(2023)\citenamefont {Tan}, \citenamefont {Qi}, \citenamefont {Chen}, \citenamefont {Danner}, \citenamefont {Kanchanawong},\ and\ \citenamefont {Tsang}}]{Tan2023}%
  \BibitemOpen
  \bibfield  {author} {\bibinfo {author} {\bibfnamefont {X.-J.}\ \bibnamefont {Tan}}, \bibinfo {author} {\bibfnamefont {L.}~\bibnamefont {Qi}}, \bibinfo {author} {\bibfnamefont {L.}~\bibnamefont {Chen}}, \bibinfo {author} {\bibfnamefont {A.~J.}\ \bibnamefont {Danner}}, \bibinfo {author} {\bibfnamefont {P.}~\bibnamefont {Kanchanawong}},\ and\ \bibinfo {author} {\bibfnamefont {M.}~\bibnamefont {Tsang}},\ }\href {https://doi.org/10.1364/OPTICA.493227} {\bibfield  {journal} {\bibinfo  {journal} {Optica}\ }\textbf {\bibinfo {volume} {10}},\ \bibinfo {pages} {1189} (\bibinfo {year} {2023})}\BibitemShut {NoStop}%
\bibitem [{\citenamefont {Datta}\ \emph {et~al.}(2020)\citenamefont {Datta}, \citenamefont {Jarzyna}, \citenamefont {Len}, \citenamefont {\L{}ukanowski}, \citenamefont {Ko\l{}ody\ifmmode~\acute{n}\else \'{n}\fi{}ski},\ and\ \citenamefont {Banaszek}}]{Datta2020}%
  \BibitemOpen
  \bibfield  {author} {\bibinfo {author} {\bibfnamefont {C.}~\bibnamefont {Datta}}, \bibinfo {author} {\bibfnamefont {M.}~\bibnamefont {Jarzyna}}, \bibinfo {author} {\bibfnamefont {Y.~L.}\ \bibnamefont {Len}}, \bibinfo {author} {\bibfnamefont {K.}~\bibnamefont {\L{}ukanowski}}, \bibinfo {author} {\bibfnamefont {J.}~\bibnamefont {Ko\l{}ody\ifmmode~\acute{n}\else \'{n}\fi{}ski}},\ and\ \bibinfo {author} {\bibfnamefont {K.}~\bibnamefont {Banaszek}},\ }\href {https://doi.org/10.1103/PhysRevA.102.063526} {\bibfield  {journal} {\bibinfo  {journal} {Phys. Rev. A}\ }\textbf {\bibinfo {volume} {102}},\ \bibinfo {pages} {063526} (\bibinfo {year} {2020})}\BibitemShut {NoStop}%
\bibitem [{\citenamefont {Datta}\ \emph {et~al.}(2021)\citenamefont {Datta}, \citenamefont {Len}, \citenamefont {Łukanowski}, \citenamefont {Banaszek},\ and\ \citenamefont {Jarzyna}}]{Datta2021}%
  \BibitemOpen
  \bibfield  {author} {\bibinfo {author} {\bibfnamefont {C.}~\bibnamefont {Datta}}, \bibinfo {author} {\bibfnamefont {Y.~L.}\ \bibnamefont {Len}}, \bibinfo {author} {\bibfnamefont {K.}~\bibnamefont {Łukanowski}}, \bibinfo {author} {\bibfnamefont {K.}~\bibnamefont {Banaszek}},\ and\ \bibinfo {author} {\bibfnamefont {M.}~\bibnamefont {Jarzyna}},\ }\href {https://doi.org/10.1364/OE.433990} {\bibfield  {journal} {\bibinfo  {journal} {Optics Express}\ }\textbf {\bibinfo {volume} {29}},\ \bibinfo {pages} {35592} (\bibinfo {year} {2021})}\BibitemShut {NoStop}%
\bibitem [{\citenamefont {Yang}\ \emph {et~al.}(2017)\citenamefont {Yang}, \citenamefont {Nair}, \citenamefont {Tsang}, \citenamefont {Simon},\ and\ \citenamefont {Lvovsky}}]{Yang2017}%
  \BibitemOpen
  \bibfield  {author} {\bibinfo {author} {\bibfnamefont {F.}~\bibnamefont {Yang}}, \bibinfo {author} {\bibfnamefont {R.}~\bibnamefont {Nair}}, \bibinfo {author} {\bibfnamefont {M.}~\bibnamefont {Tsang}}, \bibinfo {author} {\bibfnamefont {C.}~\bibnamefont {Simon}},\ and\ \bibinfo {author} {\bibfnamefont {A.~I.}\ \bibnamefont {Lvovsky}},\ }\href {https://doi.org/10.1103/PhysRevA.96.063829} {\bibfield  {journal} {\bibinfo  {journal} {Physical Review A}\ }\textbf {\bibinfo {volume} {96}},\ \bibinfo {pages} {063829} (\bibinfo {year} {2017})}\BibitemShut {NoStop}%
\bibitem [{\citenamefont {Jaffrennou}\ \emph {et~al.}(2010)\citenamefont {Jaffrennou}, \citenamefont {Claudon}, \citenamefont {Bazin}, \citenamefont {Malik}, \citenamefont {Reitzenstein}, \citenamefont {Worschech}, \citenamefont {Kamp}, \citenamefont {Forchel},\ and\ \citenamefont {Gérard}}]{Jaffrennou2010}%
  \BibitemOpen
  \bibfield  {author} {\bibinfo {author} {\bibfnamefont {P.}~\bibnamefont {Jaffrennou}}, \bibinfo {author} {\bibfnamefont {J.}~\bibnamefont {Claudon}}, \bibinfo {author} {\bibfnamefont {M.}~\bibnamefont {Bazin}}, \bibinfo {author} {\bibfnamefont {N.~S.}\ \bibnamefont {Malik}}, \bibinfo {author} {\bibfnamefont {S.}~\bibnamefont {Reitzenstein}}, \bibinfo {author} {\bibfnamefont {L.}~\bibnamefont {Worschech}}, \bibinfo {author} {\bibfnamefont {M.}~\bibnamefont {Kamp}}, \bibinfo {author} {\bibfnamefont {A.}~\bibnamefont {Forchel}},\ and\ \bibinfo {author} {\bibfnamefont {J.-M.}\ \bibnamefont {Gérard}},\ }\href {https://doi.org/10.1063/1.3315869} {\bibfield  {journal} {\bibinfo  {journal} {Applied Physics Letters}\ }\textbf {\bibinfo {volume} {96}},\ \bibinfo {pages} {071103} (\bibinfo {year} {2010})}\BibitemShut {NoStop}%
\bibitem [{\citenamefont {Andrews}\ \emph {et~al.}(2014)\citenamefont {Andrews}, \citenamefont {Peterson}, \citenamefont {Purdy}, \citenamefont {Cicak}, \citenamefont {Simmonds}, \citenamefont {Regal},\ and\ \citenamefont {Lehnert}}]{Andrews2014}%
  \BibitemOpen
  \bibfield  {author} {\bibinfo {author} {\bibfnamefont {R.~W.}\ \bibnamefont {Andrews}}, \bibinfo {author} {\bibfnamefont {R.~W.}\ \bibnamefont {Peterson}}, \bibinfo {author} {\bibfnamefont {T.~P.}\ \bibnamefont {Purdy}}, \bibinfo {author} {\bibfnamefont {K.}~\bibnamefont {Cicak}}, \bibinfo {author} {\bibfnamefont {R.~W.}\ \bibnamefont {Simmonds}}, \bibinfo {author} {\bibfnamefont {C.~A.}\ \bibnamefont {Regal}},\ and\ \bibinfo {author} {\bibfnamefont {K.~W.}\ \bibnamefont {Lehnert}},\ }\href {https://doi.org/10.1038/nphys2911} {\bibfield  {journal} {\bibinfo  {journal} {Nature Physics}\ }\textbf {\bibinfo {volume} {10}},\ \bibinfo {pages} {321} (\bibinfo {year} {2014})}\BibitemShut {NoStop}%
\bibitem [{\citenamefont {Bagci}\ \emph {et~al.}(2014)\citenamefont {Bagci}, \citenamefont {Simonsen}, \citenamefont {Schmid}, \citenamefont {Villanueva}, \citenamefont {Zeuthen}, \citenamefont {Appel}, \citenamefont {Taylor}, \citenamefont {S{\o}rensen}, \citenamefont {Usami}, \citenamefont {Schliesser},\ and\ \citenamefont {Polzik}}]{Bagci2014}%
  \BibitemOpen
  \bibfield  {author} {\bibinfo {author} {\bibfnamefont {T.}~\bibnamefont {Bagci}}, \bibinfo {author} {\bibfnamefont {A.}~\bibnamefont {Simonsen}}, \bibinfo {author} {\bibfnamefont {S.}~\bibnamefont {Schmid}}, \bibinfo {author} {\bibfnamefont {L.~G.}\ \bibnamefont {Villanueva}}, \bibinfo {author} {\bibfnamefont {E.}~\bibnamefont {Zeuthen}}, \bibinfo {author} {\bibfnamefont {J.}~\bibnamefont {Appel}}, \bibinfo {author} {\bibfnamefont {J.~M.}\ \bibnamefont {Taylor}}, \bibinfo {author} {\bibfnamefont {A.}~\bibnamefont {S{\o}rensen}}, \bibinfo {author} {\bibfnamefont {K.}~\bibnamefont {Usami}}, \bibinfo {author} {\bibfnamefont {A.}~\bibnamefont {Schliesser}},\ and\ \bibinfo {author} {\bibfnamefont {E.~S.}\ \bibnamefont {Polzik}},\ }\href {https://doi.org/10.1038/nature13029} {\bibfield  {journal} {\bibinfo  {journal} {Nature}\ }\textbf {\bibinfo {volume} {507}},\ \bibinfo {pages} {81} (\bibinfo {year} {2014})}\BibitemShut {NoStop}%
\bibitem [{\citenamefont {Thomas}\ \emph {et~al.}(2021)\citenamefont {Thomas}, \citenamefont {Parniak}, \citenamefont {Østfeldt}, \citenamefont {Møller}, \citenamefont {Bærentsen}, \citenamefont {Tsaturyan}, \citenamefont {Schliesser}, \citenamefont {Appel}, \citenamefont {Zeuthen},\ and\ \citenamefont {Polzik}}]{Thomas2021}%
  \BibitemOpen
  \bibfield  {author} {\bibinfo {author} {\bibfnamefont {R.~A.}\ \bibnamefont {Thomas}}, \bibinfo {author} {\bibfnamefont {M.}~\bibnamefont {Parniak}}, \bibinfo {author} {\bibfnamefont {C.}~\bibnamefont {Østfeldt}}, \bibinfo {author} {\bibfnamefont {C.~B.}\ \bibnamefont {Møller}}, \bibinfo {author} {\bibfnamefont {C.}~\bibnamefont {Bærentsen}}, \bibinfo {author} {\bibfnamefont {Y.}~\bibnamefont {Tsaturyan}}, \bibinfo {author} {\bibfnamefont {A.}~\bibnamefont {Schliesser}}, \bibinfo {author} {\bibfnamefont {J.}~\bibnamefont {Appel}}, \bibinfo {author} {\bibfnamefont {E.}~\bibnamefont {Zeuthen}},\ and\ \bibinfo {author} {\bibfnamefont {E.~S.}\ \bibnamefont {Polzik}},\ }\href {https://doi.org/10.1038/s41567-020-1031-5} {\bibfield  {journal} {\bibinfo  {journal} {Nature Physics}\ }\textbf {\bibinfo {volume} {17}},\ \bibinfo {pages} {228} (\bibinfo {year} {2021})}\BibitemShut {NoStop}%
\bibitem [{\citenamefont {Bor{\'o}wka}\ \emph {et~al.}(2023)\citenamefont {Bor{\'o}wka}, \citenamefont {Pylypenko}, \citenamefont {Mazelanik},\ and\ \citenamefont {Parniak}}]{Borowka2023}%
  \BibitemOpen
  \bibfield  {author} {\bibinfo {author} {\bibfnamefont {S.}~\bibnamefont {Bor{\'o}wka}}, \bibinfo {author} {\bibfnamefont {U.}~\bibnamefont {Pylypenko}}, \bibinfo {author} {\bibfnamefont {M.}~\bibnamefont {Mazelanik}},\ and\ \bibinfo {author} {\bibfnamefont {M.}~\bibnamefont {Parniak}},\ }\bibfield  {journal} {\bibinfo  {journal} {Nature Photonics}\ }\href {https://doi.org/10.1038/s41566-023-01295-w} {10.1038/s41566-023-01295-w} (\bibinfo {year} {2023})\BibitemShut {NoStop}%
\bibitem [{\citenamefont {Kumar}\ \emph {et~al.}(2023)\citenamefont {Kumar}, \citenamefont {Suleymanzade}, \citenamefont {Stone}, \citenamefont {Taneja}, \citenamefont {Anferov}, \citenamefont {Schuster},\ and\ \citenamefont {Simon}}]{Kumar2023}%
  \BibitemOpen
  \bibfield  {author} {\bibinfo {author} {\bibfnamefont {A.}~\bibnamefont {Kumar}}, \bibinfo {author} {\bibfnamefont {A.}~\bibnamefont {Suleymanzade}}, \bibinfo {author} {\bibfnamefont {M.}~\bibnamefont {Stone}}, \bibinfo {author} {\bibfnamefont {L.}~\bibnamefont {Taneja}}, \bibinfo {author} {\bibfnamefont {A.}~\bibnamefont {Anferov}}, \bibinfo {author} {\bibfnamefont {D.~I.}\ \bibnamefont {Schuster}},\ and\ \bibinfo {author} {\bibfnamefont {J.}~\bibnamefont {Simon}},\ }\href {https://doi.org/10.1038/s41586-023-05740-2} {\bibfield  {journal} {\bibinfo  {journal} {Nature}\ }\textbf {\bibinfo {volume} {615}},\ \bibinfo {pages} {614} (\bibinfo {year} {2023})}\BibitemShut {NoStop}%
\bibitem [{\citenamefont {Tu}\ \emph {et~al.}(2022)\citenamefont {Tu}, \citenamefont {Liao}, \citenamefont {Zhang}, \citenamefont {Liu}, \citenamefont {Zheng}, \citenamefont {Yang}, \citenamefont {Zhang}, \citenamefont {Yan},\ and\ \citenamefont {Zhu}}]{Tu2022}%
  \BibitemOpen
  \bibfield  {author} {\bibinfo {author} {\bibfnamefont {H.-T.}\ \bibnamefont {Tu}}, \bibinfo {author} {\bibfnamefont {K.-Y.}\ \bibnamefont {Liao}}, \bibinfo {author} {\bibfnamefont {Z.-X.}\ \bibnamefont {Zhang}}, \bibinfo {author} {\bibfnamefont {X.-H.}\ \bibnamefont {Liu}}, \bibinfo {author} {\bibfnamefont {S.-Y.}\ \bibnamefont {Zheng}}, \bibinfo {author} {\bibfnamefont {S.-Z.}\ \bibnamefont {Yang}}, \bibinfo {author} {\bibfnamefont {X.-D.}\ \bibnamefont {Zhang}}, \bibinfo {author} {\bibfnamefont {H.}~\bibnamefont {Yan}},\ and\ \bibinfo {author} {\bibfnamefont {S.-L.}\ \bibnamefont {Zhu}},\ }\href {https://doi.org/10.1038/s41566-022-00959-3} {\bibfield  {journal} {\bibinfo  {journal} {Nature Photonics}\ }\textbf {\bibinfo {volume} {16}},\ \bibinfo {pages} {291} (\bibinfo {year} {2022})}\BibitemShut {NoStop}%
\bibitem [{\citenamefont {Jing}\ \emph {et~al.}(2020)\citenamefont {Jing}, \citenamefont {Hu}, \citenamefont {Ma}, \citenamefont {Zhang}, \citenamefont {Zhang}, \citenamefont {Xiao},\ and\ \citenamefont {Jia}}]{Jing2020}%
  \BibitemOpen
  \bibfield  {author} {\bibinfo {author} {\bibfnamefont {M.}~\bibnamefont {Jing}}, \bibinfo {author} {\bibfnamefont {Y.}~\bibnamefont {Hu}}, \bibinfo {author} {\bibfnamefont {J.}~\bibnamefont {Ma}}, \bibinfo {author} {\bibfnamefont {H.}~\bibnamefont {Zhang}}, \bibinfo {author} {\bibfnamefont {L.}~\bibnamefont {Zhang}}, \bibinfo {author} {\bibfnamefont {L.}~\bibnamefont {Xiao}},\ and\ \bibinfo {author} {\bibfnamefont {S.}~\bibnamefont {Jia}},\ }\href {https://doi.org/10.1038/s41567-020-0918-5} {\bibfield  {journal} {\bibinfo  {journal} {Nature Physics}\ }\textbf {\bibinfo {volume} {16}},\ \bibinfo {pages} {911} (\bibinfo {year} {2020})}\BibitemShut {NoStop}%
\bibitem [{\citenamefont {Efron}(1979)}]{Efron1979}%
  \BibitemOpen
  \bibfield  {author} {\bibinfo {author} {\bibfnamefont {B.}~\bibnamefont {Efron}},\ }\href {https://www.jstor.org/stable/2958830} {\bibfield  {journal} {\bibinfo  {journal} {The Annals of Statistics}\ }\textbf {\bibinfo {volume} {7}},\ \bibinfo {pages} {1} (\bibinfo {year} {1979})}\BibitemShut {NoStop}%
\bibitem [{\citenamefont {Kay}(1993)}]{Kay1993}%
  \BibitemOpen
  \bibfield  {author} {\bibinfo {author} {\bibfnamefont {S.~M.}\ \bibnamefont {Kay}},\ }\href@noop {} {\emph {\bibinfo {title} {Fundamentals of Statistical Signal Processing: Estimation Theory}}},\ edited by\ \bibinfo {editor} {\bibfnamefont {S.~E. A.~V.}\ \bibnamefont {Oppenheim}}\ (\bibinfo  {publisher} {Prenitce Hall},\ \bibinfo {year} {1993})\BibitemShut {NoStop}%
\bibitem [{\citenamefont {Allevi}\ \emph {et~al.}(2013)\citenamefont {Allevi}, \citenamefont {Bondani}, \citenamefont {Marian}, \citenamefont {Marian},\ and\ \citenamefont {Olivares}}]{Allevi2013}%
  \BibitemOpen
  \bibfield  {author} {\bibinfo {author} {\bibfnamefont {A.}~\bibnamefont {Allevi}}, \bibinfo {author} {\bibfnamefont {M.}~\bibnamefont {Bondani}}, \bibinfo {author} {\bibfnamefont {P.}~\bibnamefont {Marian}}, \bibinfo {author} {\bibfnamefont {T.~A.}\ \bibnamefont {Marian}},\ and\ \bibinfo {author} {\bibfnamefont {S.}~\bibnamefont {Olivares}},\ }\href {https://doi.org/10.1364/JOSAB.30.002621} {\bibfield  {journal} {\bibinfo  {journal} {J. Opt. Soc. Am. B}\ }\textbf {\bibinfo {volume} {30}},\ \bibinfo {pages} {2621} (\bibinfo {year} {2013})}\BibitemShut {NoStop}%
\bibitem [{\citenamefont {Olivares}(2012)}]{Olivares2012}%
  \BibitemOpen
  \bibfield  {author} {\bibinfo {author} {\bibfnamefont {S.}~\bibnamefont {Olivares}},\ }\href {https://doi.org/10.1140/epjst/e2012-01532-4} {\bibfield  {journal} {\bibinfo  {journal} {The European Physical Journal Special Topics}\ }\textbf {\bibinfo {volume} {203}},\ \bibinfo {pages} {3} (\bibinfo {year} {2012})}\BibitemShut {NoStop}%
\bibitem [{\citenamefont {Tsaturyan}\ \emph {et~al.}(2017)\citenamefont {Tsaturyan}, \citenamefont {Barg}, \citenamefont {Polzik},\ and\ \citenamefont {Schliesser}}]{Tsaturyan2017}%
  \BibitemOpen
  \bibfield  {author} {\bibinfo {author} {\bibfnamefont {Y.}~\bibnamefont {Tsaturyan}}, \bibinfo {author} {\bibfnamefont {A.}~\bibnamefont {Barg}}, \bibinfo {author} {\bibfnamefont {E.~S.}\ \bibnamefont {Polzik}},\ and\ \bibinfo {author} {\bibfnamefont {A.}~\bibnamefont {Schliesser}},\ }\href {https://doi.org/10.1038/nnano.2017.101} {\bibfield  {journal} {\bibinfo  {journal} {Nature Nanotechnology}\ }\textbf {\bibinfo {volume} {12}},\ \bibinfo {pages} {776} (\bibinfo {year} {2017})}\BibitemShut {NoStop}%
\bibitem [{\citenamefont {Krokosz}\ \emph {et~al.}(2023)\citenamefont {Krokosz}, \citenamefont {Mazelanik}, \citenamefont {Lipka}, \citenamefont {Jarzyna}, \citenamefont {Wasilewski}, \citenamefont {Banaszek},\ and\ \citenamefont {Parniak}}]{dataset}%
  \BibitemOpen
  \bibfield  {author} {\bibinfo {author} {\bibfnamefont {W.}~\bibnamefont {Krokosz}}, \bibinfo {author} {\bibfnamefont {M.}~\bibnamefont {Mazelanik}}, \bibinfo {author} {\bibfnamefont {M.}~\bibnamefont {Lipka}}, \bibinfo {author} {\bibfnamefont {M.}~\bibnamefont {Jarzyna}}, \bibinfo {author} {\bibfnamefont {W.}~\bibnamefont {Wasilewski}}, \bibinfo {author} {\bibfnamefont {K.}~\bibnamefont {Banaszek}},\ and\ \bibinfo {author} {\bibfnamefont {M.}~\bibnamefont {Parniak}},\ }\href {https://doi.org/10.7910/DVN/F4LRZR} {\bibinfo {title} {{Replication Data for: Beating the spectroscopic Rayleigh limit via post-processed heterodyne detection}}} (\bibinfo {year} {2023})\BibitemShut {NoStop}%
\end{thebibliography}%

\end{document}